\documentclass[pra,twocolumn,showpacs]{revtex4}

\usepackage{graphics,amsmath,amssymb}

\newcommand{\rv}{\ensuremath{\vec{r} }}
\newcommand{\pa}{\ensuremath{\phi_{a}(x)} }
\newcommand{\pb}{\ensuremath{\phi_{b}(x)} }
\newcommand{\npa}{\ensuremath{\left|\phi_{a}\right|} }
\newcommand{\npb}{\ensuremath{\left|\phi_{b}\right|} }
\newcommand{\ha}{\ensuremath{\hat{a}} }
\newcommand{\hb}{\ensuremath{\hat{b}} }
\newcommand{\had}{\ensuremath{\hat{a}^\dagger} }
\newcommand{\hbd}{\ensuremath{\hat{b}^\dagger} }
\newcommand{\hpid}{\ensuremath{\hat{\psi}_i^\dagger} }
\newcommand{\hpi}{\ensuremath{\hat{\psi}_i} }

\newcommand{\jy}{\ensuremath{\hat{J}_y} }
\newcommand{\jz}{\ensuremath{\hat{J}_z} }
\newcommand{\sx}{\ensuremath{\hat{S}_x} }
\newcommand{\sy}{\ensuremath{\hat{S}_y} }
\newcommand{\sz}{\ensuremath{\hat{S}_z} }
\newcommand{\spl}{\ensuremath{\hat{S}_+} }
\newcommand{\sm}{\ensuremath{\hat{S}_-} }
\newcommand{\jt}{\ensuremath{\hat{J}_{\theta}} }

\newcommand{\mjx}{\ensuremath{\langle\hat{J}_x\rangle} }
\newcommand{\mjy}{\ensuremath{\langle\hat{J}_y\rangle} }
\newcommand{\mjnu}{\ensuremath{\langle\hat{J}_\nu\rangle} }

\begin{document}

\title{Positive P simulations of spin squeezing in a two-component
 Bose condensate}

\author{Uffe V. Poulsen}
\email{uvp@ifa.au.dk}
\author{Klaus M{\o}lmer}
\affiliation{ 
Institute of Physics and Astronomy, University of Aarhus,
    DK-8000 \AA rhus C, Denmark}


\begin{abstract}
  The collisional interaction in a Bose condensate represents a
  non-linearity which in analogy with non-linear optics gives rise to
  unique quantum features. In this paper we apply a Monte Carlo method
  based on the positive P pseudo-probability distribution from quantum
  optics to analyze the efficiency of spin squeezing by collisions in
  a two-component condensate. The squeezing can be controlled by
  choosing appropiate collision parameters or by manipulating the
  motional states of the two components.
\end{abstract}  

\maketitle

\section{Introduction}
\label{sec:intro}

In quantum optics it was realized some time ago that so-called
'non-classical' states of light, in particular, the so-called squeezed
states, may outperform classical field states in high precision
experiments~\cite{fabre92:_quant_noise_reduc_optic_system}. Moderately
squeezed light has been produced, and several non-classical properties
have been demonstrated, but classical technology has so far left
enough space for improvements that a real practical high precision
application of squeezed light remains to be seen.  Atomic squeezing is
defined in a way similar to squeezing of light, in the sense that the
quantum mechanical uncertainty of a physical variable, here a
component of the collective spin, is reduced
by a suitable manipulation of the quantum state of the system.\\
Due to quite strong interactions between atoms and between atoms and
light, and due to the long storage and interaction times for atoms,
the degree of correlations and squeezing obtainable in atomic systems
exceeds the one in light by orders of magnitude.  Also, there is a
real potential for practical application of squeezed atoms, since
current atom interferometers and atomic clocks operate at a limit of
precision which can only be improved by imposing quantum correlations
between the atoms.\\
Recently there has been a number of proposals for practical spin
squeezing: absorption of squeezed light~\cite{kuzmich97}, quantum
non-demolition atomic detection~\cite{molmer99}, collisional
interactions in classical or degenerate gasses~\cite{pu00,
  sorensen01:_many_bose}, photo-dissociation of molecular
condensates~\cite{poulsen01:_quant_states_bose_einst}, and controlled
dynamics in quantum computers with ions or
atoms~\cite{molmersorensen}.  Since quantum mechanical squeezing
implies quantum mechanical uncertainties below the level in 'natural'
states of the system, {\it e.g.}, the ground state of the systems, the
analysis of squeezing has to be very precise, and in particular one
should avoid use of classical approximations or assumptions. In the
present paper we shall investigate the possibilities for squeezing in
a Bose-Einstein condensate, using a method which takes the
interactions and the multi-mode character of the problem exactly into
account. We confirm the validity of results of a recent
study~\cite{sorensen01:_many_bose}, and we propose an alternative
scheme for spin squeezing which relies on a spatial
separation of atoms in different internal states.

We consider in this paper a Bose-Einstein condensate of two-level
atoms in a trap. The dynamics of such a system is in the second
quantized formalism with creation and annihilation operators of atoms
in the state $i=a$ or $b$, \hpid, \hpi, controlled by the Hamiltonian:
\begin{multline}
  \label{eq:def_H}
    \hat{H} =\\ 
    \int \! d^3 r \Bigl\{
     \sum_{i=a,b} \left[
      \hat{\psi}^{\dagger}_i(\vec{r}) \hat h_i
      \hat{\psi}(\vec{r}) %\\
      + \frac{g_{ii}}{2}
      \hat{\psi}^{\dagger}_i(\vec{r})
      \hat{\psi}^{\dagger}_i(\vec{r}) 
      \hat{\psi}_i(\vec{r}) \hat{\psi}_i(\vec{r}) 
    \right]
    \\
    + g_{ab}       
    \hat{\psi}^{\dagger}_a(\vec{r})
    \hat{\psi}^{\dagger}_b(\vec{r}) 
    \hat{\psi}_b(\vec{r}) \hat{\psi}_a(\vec{r}) 
    \Bigl\}
    .
\end{multline}
Here, $h_i$ is the single particle Hamiltonian for atoms in internal state
$i$ and $g_{ij}$ is the effective two-body interaction strength
between an atom in state $i$ and one in state $j$. In
terms of the corresponding scattering lengths they are given by
$g_{ij}=4\pi\hbar^2 a_{ij} / m $. 

At temperatures sufficiently below the critical temperature for
Bose-Einstein condensation almost all atoms occupy the same single
particle wavefunction which to a very good approximation can be
obtained from a two-component Gross-Pitaevskii Equation.  There are,
however, important effects about which the Gross-Pitaevskii Equation
gives no information: the population statistics of the condensate is
assumed to be poissonian (or a number
state~\cite{castin98:_low_bose_einst}), it is not treated as a
variable which has to be determined, and which can be manipulated by
physical processes. In multi-component condensates, the relative
populations and coherences of different states are important degrees
of freedom which require a more elaborate treatment.

\section{Two-mode model}
\label{sec:twomode}
 
In a first attempt to model the population statistics it is convenient
for simplicity to assume a separation of the spatial and
internal degrees of freedom
\begin{equation}
  \label{eq:sep_spa_int}
  \hat\psi_a(\rv) = \hat{a} \phi_a(\rv) 
  \qquad  
  \hat\psi_b(\rv) = \hat{b} \phi_b(\rv)
  .
\end{equation}
The separation (\ref{eq:sep_spa_int}) is difficult to justify in general
but it is certainly reasonable in the initial state that we have in
mind: A very pure (T=0) single component condensate is prepared. We
then apply a $\pi/2$-pulse, coherently transfering all atoms to an
equal superposition of $a$ and $b$. Then Eq.(\ref{eq:sep_spa_int}) is
fulfilled to a good approximation with $\phi_a=\phi_b$. 

In the subsequent dynamics, the wavefunctions \pa and \pb may evolve
with time, and we describe this evolution with Gross-Pitaevskii Equations
\begin{equation}
  \label{eq:gpe}
  i \hbar\partial_t \phi_i =
  \left( \hat{h}_i + g_{ii} \frac{N}{2} \left|\phi_i\right|^2 
    + g_{i\underline{i}} \frac{N}{2} \left|
      \phi_{\underline{i}} \right|^2 
  \right) \phi_i
\end{equation}
where  $\underline{a} \equiv b$ and {\it vice versa}. 

The dynamics associated with distribution of atoms among the two modes
is now studied by rewriting Eq.(\ref{eq:def_H}) in terms of the \ha
and \hb operators:
\begin{multline}
  \label{eq:H_twomode}
   \hat{H}_{\text{two-mode}} =
  g_{bb} \left( \hbd\hbd\hb\hb - \frac{N}{2}\hbd\hb \right) 
  \int dr^3 \npb^4 \\
  +g_{aa}  \left( \had\had\ha\ha - \frac{N}{2}\had\ha \right) 
  \int dr^3 \npa^4 \\
  +g_{ab} \left( \hbd\had\ha\hb - \frac{N}{2}\hbd\hb - \frac{N}{2}\had\ha \right) 
  \int dr^3 \npb^2\npa^2
  .
\end{multline}
All terms in this Hamiltonian commute with $\had\ha$ and $\hbd\hb$
and so the interesting dynamics takes place in the coherences
between the two internal states. To study these coherences it is
convenient to define effective internal spin operators
\begin{equation}
  \label{eq:def_int_spin}
  \hat{S}_+ \equiv \hat{b}^\dagger\hat{a} 
  \quad  
  \hat{S}_- \equiv \hat{a}^\dagger\hat{b}
  \quad
    \hat{S}_z \equiv \frac{1}{2}\left(
      \hat{b}^\dagger\hat{b} 
      - \hat{a}^\dagger\hat{a} 
    \right).
\end{equation}
$\hat{S}_z$ represents the difference of total population of states
$a$ and $b$, a quantity which can be measured, {\it e.g.}, by laser
induced fluoresence in an experiment. $\hat{S}_+$ and $\hat{S}_-$ or,
equivalently, $\sx \equiv (\spl + \sm)/2$ and $\sy \equiv (\spl -
\sm)/2i$ represent the coherences between $a$ and $b$.  By suitable
couplings they can also be turned into population differences and are
therefore interesting observables of the system.  In fact it turns out
that the quantity that determines the accuracy of a given
spectroscopic experiment is the ratio of the measured spin-component
(the signal) to the fluctuations in a component perpendicular to it
(the noise)~\cite{wineland94:_squeez}. For the initial state prepared
by a $\pi/2$ pulse we have $N$ independent atoms all in the same equal
superposition of $a$ and $b$. Then the mean spin is in the $xy$-plane
and by definition we can take it to be along the $x$-axis. The maximal
signal we can obtain is the length of the spin $S=N/2$ and the
perpendicular fluctuations are then $S/2$. To improve this ``standard
quantum limit'' signal/noise ratio we can introduce correlations among
the atoms {\it i.e.} we can introduce spin squeezing.

With the internal spin operators of Eq.~(\ref{eq:def_int_spin})
the Hamiltonian (\ref{eq:H_twomode}) can be written
\begin{multline}
  \label{eq:twomode_H}
  \hat{H}_{\text{two-mode}} = e(t) \hat{N} %\\
  +  k(t) \hat{S}_z \\
  + E(t) \hat{N}^2 %\\
  + D(t) \hat{N}\hat{S}_z
  + \chi(t) \hat{S}_z^2
\end{multline}
where the time dependent coefficients are given by integrals involving
the time dependent mode functions found from Eq.(\ref{eq:gpe}).  All
the terms in the Hamiltonian (\ref{eq:twomode_H}) commute which is an
important simplification as the coefficients are time-dependent. The
term proportional to $\hat{S}_z$ will result in a rotation of the spin
around the $z$-axis. The $\hat{N}$ and $\hat{N}^2$ terms give
different dynamical overall phases to states with different total
numbers of atoms. Such phases are immaterial if we have no external
phase-standard to compare with and we neglect these terms for the
purpose of the present work. The $\hat{N}\hat{S}_z$-term adds to the
spin rotation with an angle linear in $N$. It cannot be neglected as
the direction of the spin (the phase between $a$- and $b$-components)
can be probed by a second $\pi/2$ pulse phase locked to the first
$\pi/2$ pulse. The term of interest to us is $\hat{S}_z^2$, the effect
of which is well known from the work of Kitagawa and
Ueda~\cite{kitagawa93:_squeez}. It produces spin-squeezing, {\it i.e.}
it entangles the individual atoms in a way that reduces the
fluctuations of the total spin in one of the directions perpendicular
to the average spin. The strength parameter of the squeezing operator
is given by
\begin{equation}
  \label{eq:def_chi}
  \chi(t)  = \int \! dr^3 \;
  \frac{1}{2}
  \left(
    g_{bb}\left|\phi_b\right|^4
    +g_{aa}\left|\phi_a\right|^4
    -2g_{ab}\left|\phi_a\right|^2\left|\phi_b\right|^2
  \right)
  .
\end{equation}
Given the time integral $\mu = 2 \int\!\chi(t') dt'$ of this parameter
Kitagawa and Ueda provide the analytical expressions for the variance
of the squeezed spin component
\begin{multline}
  \label{eq:ueda_ds2}
  \langle \Delta S_\theta^2 \rangle
  = \\
  \frac{S}{2}\left\{[1+\frac{1}{2}(S-\frac{1}{2})A]
    -\frac{1}{2}(S-\frac{1}{2})
    \sqrt{A^2+B^2}]\right\}
\end{multline}
where $A=1-\cos^{2S-1}\mu$ and $B=4\sin\mu/2\cos^{2S-2}\mu/2$, and
they specify the direction of the squeezed spin component
$\hat{S}_\theta=\cos\theta\sy+\sin\theta\sz$:
\begin{equation}
  \label{eq:ueda_theta}
  \theta=\frac{\pi}{2}+\frac{1}{2}\arctan\frac{B}{A}.
\end{equation}
It is of some interest to use simple
analytical approximations for $\chi(t)$ and we will do so in the
specific cases studied below. Another approach would of course be to
obtain $\chi(t)$ by a numerical solution of Eq.~(\ref{eq:gpe}).

\section{Full multi-mode description}
\label{sec:multimode}

Equations (\ref{eq:sep_spa_int})-(\ref{eq:def_chi}) are based on a
simplifying assumption. The actual observables of the system are more
complicated to deal with, but we can define a set of operators obeying
angular momentum commutation relations by
\begin{align}
  \label{eq:def_j}
  \hat{J}_x &\equiv \frac{1}{2} \int dr^3 \left(
    \hat{\psi}_b^\dagger \hat{\psi}_a + \hat{\psi}_a^\dagger \hat{\psi}_b
  \right)\\
  \hat{J}_y &\equiv \frac{1}{2i} \int dr^3 \left( 
    \hat{\psi}_b^\dagger \hat{\psi}_a - \hat{\psi}_a^\dagger \hat{\psi}_b 
  \right)\\
  \hat{J}_z &\equiv \frac{1}{2} \int dr^3 \left( 
    \hat{\psi}_b^\dagger \hat{\psi}_b - \hat{\psi}_a^\dagger \hat{\psi}_a 
  \right)
  .
\end{align}
The total number operator $\hat{N}\equiv \int dr^3 (\hat{\psi}_a^\dagger
\hat{\psi}_a + \hat{\psi}_b^\dagger \hat{\psi}_b)$ commutes with these
three operators and when the two-mode approximation applies well, the
two-mode and multi-mode operators are comparable by the replacement
\begin{equation}
  \label{eq:trans_s2j}
  \hat{J}_+ = 
  \rho e^{i\nu}\hat{S}_+
  \quad
  \hat{J}_- =
  \rho e^{-i\nu}\hat{S}_-
  \quad
  \hat{J}_z = \hat{S}_z
\end{equation}
where $\rho e^{i\nu} \equiv \int dr^3 \phi_b^*(\rv)\phi_a(\rv)$. The factor
$\rho$ takes into account that \mjx and \mjy vanish unless
the atoms in state $a$ and $b$ occupy the same region in phase space
and $\nu$ is a dynamical phase from the spatial dynamics.

The two-mode approximation provides an intuitive picture of
the evolution of the system. It is however not easy to justify the
factorization (\ref{eq:sep_spa_int}) and we shall therefore apply an
exact method to determine more precisely what happens to the mean
values and the variances of the components of $\hat{\vec{J}}$. The
positive P function ($P_+$)~\cite{gardiner91:_quant_noise} is a
pseudo-probability distribution giving expectation values of normally
ordered operator products as c-number averages. In our case with two
internal states one has
\begin{equation}
  \label{eq:def_posp}
  \langle 
  :\!f[
  \hat{\boldsymbol{\psi}}(t)
  ]\!:
  \rangle =%\\ 
  \int \! d [ \boldsymbol{\psi} ]
    f[
    \boldsymbol{\psi}
    ]
  P_+ [ 
  \boldsymbol{\psi},t
  ]  
\end{equation}
where

\begin{equation}
  \label{eq:def_vecpsi}
  \hat{\boldsymbol{\psi}}(t) \equiv 
  \left(
    \begin{array}{c}
      \hat\psi_a(t) \\
      \hat\psi_a^\dagger(t) \\
      \hat\psi_b(t) \\
      \hat\psi_b^\dagger(t)
    \end{array}
  \right)
  \quad\text{and}\quad
  \boldsymbol{\psi} \equiv 
  \left(
    \begin{array}{c}
      \psi_{a1} \\
      \psi_{a2} \\
      \psi_{b1} \\
      \psi_{b2}
    \end{array}
  \right).
\end{equation}
The distribution is not determined uniquely but one particular choice
obeys a functional Fokker-Planck equation which is of course immensely
difficult to solve. For numerical purposes it is much better to
translate it to coupled Langevin equations for the 4 c-number fields
(``wave functions''). These equations are ``noisy Gross-Pitaevskii equations''
\begin{equation}
  \label{eq:langevin}
   d\psi_{i\mu}= (-1)^\mu \frac{i dt}{\hbar} \left(
    \hat{h}_i + g_{ii} \psi_{i2}\psi_{i1} 
    + g_{ab} \psi_{\underline{i}2}\psi_{\underline{i}1} 
  \right) \psi_{i\mu} + dW_{i\mu}
\end{equation}
where $\underline{a} \equiv b$ and {\it vice versa}. In order to treat
the interactions exactly (within the approximation given by the form
of Eq.(\ref{eq:def_H})) the noise terms have to be gaussian
and to fulfill:
\begin{gather}
  \label{eq:noise}
  \langle dW_{i\mu}(\rv,t) \rangle  
  = 0 \\
  \begin{split}
  \langle dW_{i\mu}(\rv,t) dW_{j\nu}&(\rv',t') \rangle  
  = \delta^{(3)}(\rv\!-\!\rv')\delta(t\!-\!t')\delta_{\mu\nu} \\ 
  &
  \times
     (-1)^\mu
  \frac{i dt g_{ij}}{\hbar} 
  \psi_{i\mu}(\rv,t)\psi_{j_\mu}(\rv,t)
  .
  \end{split}
  \end{gather}
On the computer we can simulate the Langevin equations to obtain an
ensemble of realizations of $\boldsymbol{\psi}$. This ensemble is a
finite sampling of $P_+$ and can therefore be used to calculate
expectation values {\it via} the prescription (\ref{eq:def_posp}) and
with a precision limited only by the number of realizations in the
ensemble.

Three limitations of the method should be noted: ({\it{i}}) The
initial state of the system has to be expressed as an initial $P_+$.
This is trivial for a single coherent state and for any mixture of
coherent states but it can be complicated for other initial
conditions. ({\it ii}) The $P_+$ method has a notorious divergence
problem at large non-linearities. This problem sets in after a certain
time and is clearly noticable in the simulations. We can therefore
easily tell how long we can trust the results of the method. ({\it
  iii}) Although we have so far written all equations in 3D it would
be computationally very heavy to simulate a sufficient number of
realizations of Eq.~(\ref{eq:langevin}). In what follows we will thus
restrict ourselves to 1D. It is reasonable to assume that this may
alter the quantitative results significantly but a 1D calculation can
be used to investigate the validity of the two-mode model which may
hereafter be applied in 3D with more confidence.

\section{Spin squeezing with controlled collision strengths}
\label{sec:ss_con_g}
Let us first focus on a situation with simple spatial dynamics. If we
set $V_a=V_b=m\omega^2x^2/2$ (1D model) and we assume that the values
of the collision strengths can be controlled so that, e.g.,
$g_{aa}=g_{bb}=2g_{ab}\equiv g$ the spatial dynamics is limited to a
slight breathing. The spin-dynamics is almost a pure squeezing, that
is, the mean spin stays in the $x$-direction. To get an estimate of
the strength parameter $\chi$ of Eq.~(\ref{eq:def_chi}) we find
$\phi_a(t=0)=\phi_b(t=0)$ as the Thomas-Fermi approximation to the
stationary solution of the GPE with all atoms in the $a$-state. We
then have:
\begin{equation}
  \label{eq:esti_chi}
  \chi \cong  \frac{1}{2^{2/3}3^{2/3}5}
  \frac{g^{2/3}}{N^{1/3}} m^{1/3}\omega^{2/3} 
  .
\end{equation}
Choosing $g=0.005 \hbar\omega a_0$ and $N=2000$ we get $\chi=6.1\times10^{-4}\hbar\omega$ which
should give a sizable squeezing within a quarter of a trapping
periode. $a_0\equiv\sqrt{\frac{\hbar}{m\omega}}$ is the harmonic oscillator
length in the trap and $m$ is the atomic mass.

\begin{figure}[tbp]
  \begin{center}
    \resizebox{0.5\textwidth}{!}
    {
      \includegraphics{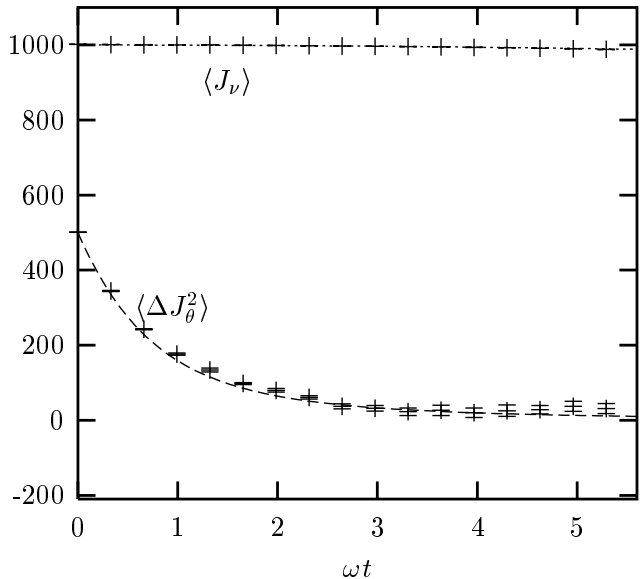}
      }
    \caption{Squeezing of the collective spin for favorable interaction
      parameters: N=2000 atoms and
      $(g_{aa},g_{ab},g_{bb})=(1.0,0.5,1.0)\times 5 \times
      10^{-3}\hbar\omega a_0 $ (1D model). $+$ (with
      errorbars) show $P_+$ results,
      lines show results of the two-mode model.}
    \label{fig:favorpara}
  \end{center}
\end{figure}

In Fig.\ref{fig:favorpara} 
we show results of both the two-mode approximation (\ref{eq:ueda_ds2})
and of the $P_+$ simulation for the parameters mentioned above. $\chi$
was assumed to be constant and of the value determined by
Eq.~(\ref{eq:esti_chi}).  $\jt=\cos\theta\jy+\sin\theta\jz$ refers to
the squeezed component of the spin. The direction $\theta$ in the
$yz$-plane is determined in the two mode model, i.e., from
Eq.~(\ref{eq:ueda_theta}) and it is not independently optimized for the
full $P_+$ results. The agreement is seen to be surprisingly good
considering the crudeness of the estimate of the parameters in the two
mode model.  Within the uncertainty of the positive P results (the
$\langle \Delta \jt^2\rangle$ data are presented with errorbars in
Fig.~\ref{fig:favorpara}) the noise suppression is seen to be almost
perfect. These calculations thus confirm the results of
S{\o}rensen {\it et al.}~\cite{sorensen01:_many_bose} where the simple
two-mode approach was supplemented by another approximate method.

\section{Spin squeezing with controlled mode functions overlaps}
\label{sec:alt_overlaps}

In the experiments on the $|F=1,m_f=-1\rangle$ and $|F=2,m_f=1\rangle$
states in $^{87}$Rb~\cite{hall98:_dynam_compo_sepa} the scattering
lengths and thus the interaction strengths are actually in proportion
$g_{aa}:g_{ab}:g_{bb}=1.03:1:0.97$. This is far from ideal conditions
for squeezing as we can see from Eq.~(\ref{eq:def_chi}): $\chi=0$ when
in addition $\phi_a=\phi_b$ as is the case initially. To produce a
sizable squeezing effect we propose to make the two mode functions
differ (see also~\cite{goldstein00:_elimin_mean_field_shift_two}).
This is achieved by applying different potentials to the two internal
states which has actually already been done for magnetically trapped
Rb making use of gravity and different magnetic moments of the two
internal states~\cite{hall98:_dynam_compo_sepa}. If $V_b$ is displaced
from $V_a$ the $b$-component created by the initial $\pi/2$-pulse will
move away from the $a$ component thereby reducing the overlap $\int\!
dx \left|\phi_a\right|^2 \left|\phi_b\right|^2$ and increasing $\chi$.
It is remarkable that the squeezing then takes place while the two
components are away from each other and are therefore not interacting.

To demonstrate the accomplishments of the scheme described above we
have chosen simply to displace $V_b$ by a certain amount $x_0$ from
$V_a$. In this model both potentials are still harmonic and of the
same strength. The spatial dynamics is now more complicated but it
can be approximated by the solutions to coupled
Gross-Pitaevskii equations.  To get a rough idea of the evolution we
can use the well know evolution of a displaced ground state
wavefunction in a harmonic trap.  We then get
\begin{equation}
  \label{eq:esti_rho_nu}
  \rho e^{i\nu} = 
   e^{ -x_0^2\left(1-\cos(\omega t) \right)/2}
   e^{ i x_0^2 \sin(\omega t) / 2}
\end{equation}
and
\begin{equation}
  \label{eq:esti_overlap}
  \int \npb^2 \npa^2 dx = \frac{1}{\sqrt{2\pi}} 
  \; e^{-x_0^2\left(1-\cos(\omega t)\right)/4}
  .
\end{equation}
If we choose $(g_{aa},g_{ab},g_{bb})=(1.03,1,0.97)\times 5 \times
10^{-3}\hbar\omega a_0 $ and $x_0=3 a_0$ this model gives an order of
magnitude estimate for the integrated strength parameter of
$\int_0^{2\pi/\omega} \chi(t) dt \cong 10^{-2}\hbar\omega$ when the
two components are again overlapped. For 2000 atoms this corresponds
to a reduction of the uncertainty in the squeezed spin component by
roughly a factor of 10 according to
Eq.~\ref{eq:ueda_ds2}.

\begin{figure}[tbp]
  \begin{center}
    \resizebox{0.5\textwidth}{!}
    {
      \includegraphics{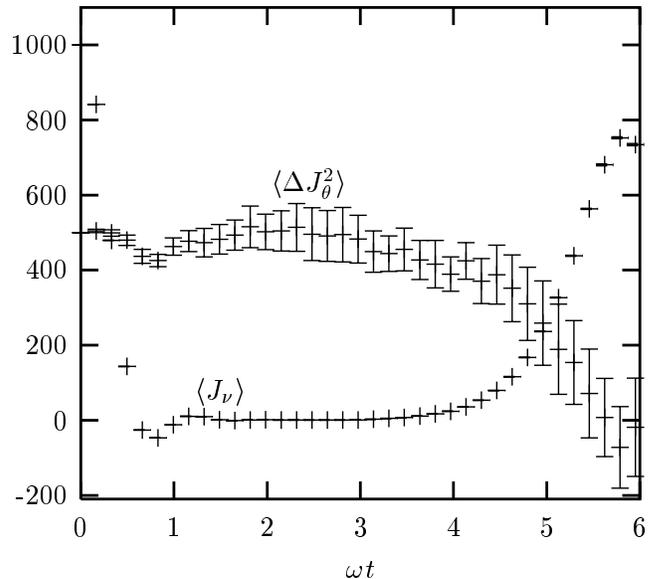}
      }
    \caption{Squeezing due to spatial separation of the two internal
      state potentials by $x_0 = 3 a_0$. The total number of atoms is
      2000 and the interaction strengths are
      $(g_{aa},g_{ab},g_{bb})=(1.03,1,0.97)\times 5 \times
      10^{-3}\hbar\omega a_0 $ (1D model).}
    \label{fig:move}
  \end{center}
\end{figure}

In Fig.\ref{fig:move}
we use these estimates to analyze the $P_+$ simulations, i.e., we plot
the expectation value of the spin component predicted to be maximal,
\mjnu, and the variance of the perpendicular component predicted to be
squeezed, $\langle\Delta \jt^2\rangle$. As can be seen, after a fast
drop a large fraction of the original mean spin is recovered in \mjnu
when the two wave packets are again overlapped. This confirms our
prediction of $\nu$ and it implies that a sizable signal can be
obtained in an experiment. At the same time the variance in the
predicted perpendicular component is strongly suppressed confirming
our prediction for $\theta$ and implying that the noise in the
experiment can be significantly reduced below the standard quantum
limit.

\section{Discussion}
\label{sec:summary}
We have presented the first exact numerical studies of spin squeezing
in two-component condensates. As in our previous work on squezing of a
single component atomic field
operator~\cite{poulsen01:_quant_states_bose_einst} we have applied the
positive P distribution which is well suited for studies of transient,
short time behavior of interacting many-body systems. This formulation
involves simulations, and due to the formal separation of c-number
representations of creation and annihilation operators, 'non-physical'
results may occur such as negative/complex atomic densities, and few
instants with negative variances as depicted in
Figs.~\ref{fig:favorpara} and \ref{fig:move}. For long times we cannot
exclude that slight imprecision due to discretization of space and
time contributes to these results. The Monte Carlo nature of the
method makes it somewhat tedious to check rigorously for
discretization errors but for resonable numbers of realizations in the
ensemble sampling errors will dominate anyhow.

The conclusion of this paper is that the predictions of the simple
two-mode model reproduce the full multimode result also
quantitatively.  This means that the spatio-temporal dynamics and the
population dynamics couple the way they should to produce spin
squeezing but the resulting entanglement between spatial and internal
degrees of freedom is small enough that purely internal state
observables show strong squeezing and multi-particle entanglement. In
particular the calculations of Fig.~\ref{fig:move} reveal that quite
significant distortions of the distributions when the components
separate and merge do not prevent sizable spin squeezing.

Although we claim quantitative agreement above it should of course be
realized that this is within the 1D model. To use the two-mode model
to describe a real experiment the spatial dynamics which enters the
internal dynamics via Eqs.~(\ref{eq:def_chi}) and (\ref{eq:trans_s2j})
must be treated with some accuracy. Apart from the 3D aspects it
should be noted that e.g. in some experiments on two-component
condensates~\cite{myatt97:_produc_overlap} the
displacement of the potentials are accompanied (and in fact due to)
different strenghts of the potentials. This adds to the complexity of
the dynamics and will of course be important if the aim is precise
quantitative predictions and not a proof-of-principle analysis as
offered in this paper.

More tests need to be carried out for other kinds of processes, but
the present study suggests that the simple two-mode description
provides good predictions for the many-body dynamics of spin
squeezing. Hence other ideas, e.g., for reducing the fluctuations in
the total number of atoms in a condensate or in out-coupled atom laser
beams, may be reliably based on the terms of Eq.~(\ref{eq:twomode_H}).

\end{document}